\documentclass[11pt,a4paper]{article}

\usepackage{jheppub}
\usepackage{amsmath}
\usepackage{amssymb}
\usepackage{amsfonts}
\usepackage{braket}
\usepackage[normalem]{ulem}
\usepackage{color}
\usepackage{bbold}
\usepackage{ulem}

\DeclareMathOperator{\tr}{tr}

\title{Holographic Renyi Entropy from Quantum Error Correction}

\author{Chris Akers}
\author{and Pratik Rath}

\affiliation{Center for Theoretical Physics and Department of Physics,\\
University of California, Berkeley, CA 94720, U.S.A. and}
\affiliation{Lawrence Berkeley National Laboratory, Berkeley, CA 94720, U.S.A.} 
\emailAdd{cakers@berkeley.edu}
\emailAdd{pratik\_rath@berkeley.edu}

\abstract{
We study Renyi entropies $S_n$ in quantum error correcting codes and compare the answer to the cosmic brane prescription for computing $\widetilde{S}_n \equiv n^2 \partial_n (\frac{n-1}{n} S_n)$. 
We find that general operator algebra codes have a similar, more general prescription.
Notably, for the AdS/CFT code to match the specific cosmic brane prescription, the code must have maximal entanglement within eigenspaces of the area operator. This gives us an improved definition of the area operator, and establishes a stronger connection between the Ryu-Takayanagi area term and the edge modes in lattice gauge theory.
We also propose a new interpretation of existing holographic tensor networks as area eigenstates instead of smooth geometries.
This interpretation would explain why tensor networks have historically had trouble modeling the Renyi entropy spectrum of holographic CFTs, and it suggests a method to construct holographic networks with the correct spectrum.
}

\begin{document}
\maketitle
\section{Introduction}\label{sec:Intro}
While the quantum error-correction interpretation of AdS/CFT was discovered by trying to resolve certain paradoxes \cite{Almheiri:2014lwa}, it has exceeded its original purpose.
Among other things, it has led to proofs of entanglement wedge reconstruction \cite{Dong:2016eik,Cotler:2017erl} and a better understanding of the black hole interior \cite{Hayden:2018khn, Almheiri:2018xdw}.

One remarkable result was an appreciation of the Ryu-Takayanagi (RT) formula  \cite{Ryu:2006bv,Ryu:2006ef,Hubeny:2007xt} as a property of the code \cite{Harlow:2016vwg}. 
The RT formula computes the von Neumann entropy $S(\rho_A) \equiv -\tr \rho_A \log \rho_A$ of a subregion $A$ of a holographic CFT via the area $\mathcal{A}$ of an extremal surface in the AdS dual:
\begin{align}\label{eqn:RT}
    S(\rho) = \frac{\mathcal{A}}{4G_N} + S_\text{bulk}
\end{align}
where $S_\text{bulk}$ is the entropy of the matter in the bulk subregion dual to $A$ \cite{Faulkner:2013ana}. 
While the area term is in general $\mathcal{O}(1/G_N)$, $S_\text{bulk}$ is in general $\mathcal{O}(1)$ and is the quantum (or ``FLM'') correction to RT \cite{Faulkner:2013ana}.

The RT formula naturally appears when one computes the von Neumann entropy of an encoded state in a quantum error-correcting code. 
We demonstrate this now in the context of the type of code we will be using throughout: an operator-algebra quantum error-correcting (OQEC) code with complementary recovery.
These appear to be the best codes to model AdS/CFT \cite{Harlow:2016vwg}, and we will explain their details in Section~\ref{sec:OQEC}.
For now, it suffices to say that we consider a finite-dimensional Hilbert space $\mathcal{H} = \mathcal{H}_A \otimes \mathcal{H}_{\bar{A}}$, and a Hilbert space $\mathcal{H}_\text{code} \subseteq \mathcal{H}$.
Furthermore, these Hilbert spaces have the following decompositions
\begin{align}
    &\mathcal{H}_A = \oplus_\alpha (\mathcal{H}_{A_1^\alpha} \otimes \mathcal{H}_{A_2^\alpha}) \oplus \mathcal{H}_{A_3}~, \\
    &\mathcal{H}_\text{code} = \oplus_\alpha (\mathcal{H}_{a_\alpha} \otimes \mathcal{H}_{\bar{a}_\alpha})~,
\end{align}
with $\dim \mathcal{H}_{A_1^\alpha} = \dim \mathcal{H}_{a_\alpha}$.
In these codes, a ``logical'' density matrix $\rho$ acting on $\mathcal{H}_\text{code}$ is encoded in a ``physical'' density matrix acting on $\mathcal{H}$.
Moreover, we say that $A$ encodes $a$ if there exists a unitary operation $U_A$ on $\mathcal{H}_A$ such that
\begin{align}\label{eqn:advancedqeccdensitymatrix}
    \widetilde{\rho}_A = U_A \left( \oplus_\alpha (p_{\alpha}\rho_{\alpha} \otimes \chi_{\alpha}) \right) U_A^\dagger~,
\end{align}
and the density matrices $\rho_{\alpha}$ act on $\mathcal{H}_{A_1^\alpha}$ and correspond to the state $\rho_{a_\alpha}$ that we wished to encode in $\widetilde{\rho}_A$.
I.e. $\rho_{\alpha}$ acts on $\mathcal{H}_{A_1^\alpha}$ in the same way that $\rho_{a_\alpha}$ does on $\mathcal{H}_{a_\alpha}$.
The density matrices $\chi_{\alpha}$ act on $\mathcal{H}_{A_2^\alpha}$ and correspond to the extra degrees of freedom that help encode.
The von Neumann entropy is
\begin{align}\label{eqn:OQECentropy}
    S(\widetilde{\rho}_A) = \tr \left( \widetilde{\rho}_A \mathcal{L}_A \right) - \sum_\alpha p_\alpha \log p_\alpha + \sum_\alpha p_\alpha S(\rho_{\alpha})
\end{align}
where the ``area operator'' is defined as 
\begin{align}
    \mathcal{L}_A \equiv \oplus_\alpha S(\chi_{\alpha}) \mathbb{1}_{a_{\alpha} \bar{a}_{\alpha}}~.
\end{align}
Compare this to Eq.~(\ref{eqn:RT}). 
The first term on the RHS of both are the expectation value of linear operators evaluated in the state of interest. 
The other terms are naturally the algebraic von Neumann entropy of the logical state \cite{Harlow:2016vwg}.
This is the basic connection between RT and quantum error-correcting codes.

In AdS/CFT, there is a natural generalization of the RT formula that we now describe.
The von Neumann entropy $S(\rho)$ has a well-known generalization called the Renyi entropies, $S_n (\rho) \equiv \frac{1}{1-n}\log \tr \rho^n$.
While the Renyi entropy equals the von Neumann entropy for $n = 1$, it is widely believed that the Renyi entropy does not satisfy anything qualitatively similar to the RT formula for $n \neq 1$ \cite{Dong:2016fnf,Almheiri:2016blp}.
However, a related quantity called the refined Renyi entropy 
\begin{align}
    \widetilde{S}_n(\rho) \equiv n^2 \partial_n \left( \frac{n-1}{n} S_n(\rho) \right)
\end{align}
also reduces to the von Neumann entropy for $n = 1$ but in fact does satisfy a generalized version of the RT formula \cite{Dong:2016fnf}.
One computes $\widetilde{S}_n(\rho_A)$ holographically via the so-called ``cosmic brane prescription,'' which works roughly as follows.
Into the state $\rho_A$, insert an extremal codimension-2 cosmic brane in AdS homologous to the boundary region $A$, and let this brane have tension $T_n = \frac{n-1}{4 n G_N}$.
The refined Renyi entropy is related to the area of this brane:
\begin{align}
    \widetilde{S}_n(\rho_A) = \frac{\mathcal{A}_\text{brane}}{4 G_N} + \widetilde{S}_{n,\text{bulk}}~,
\end{align}
where $\widetilde{S}_{n,\text{bulk}}$ is the refined Renyi entropy of matter fields in the bulk region dual to $A$.
We describe this prescription in detail in Section~\ref{sec:qecrenyi}.
As one might expect, the cosmic brane prescription limits to the RT formula as $n \to 1$, because in that limit the tension vanishes and the cosmic brane reduces to an extremal surface.

Because the cosmic brane prescription generalizes RT, it behooves us to investigate whether the connection between RT and quantum error-correcting codes can be generalized to the cosmic brane prescription. 
Let us formulate this question more precisely.
Does the refined Renyi entropy of $\widetilde{\rho}_A$ from Eq.~(\ref{eqn:advancedqeccdensitymatrix}) satisfy some formula like
\begin{align}\label{eqn:idealbranegeneralization}
    \widetilde{S}_n(\widetilde{\rho}_A) \stackrel{?}{=} \tr(\widetilde{\rho}_{\text{brane},A} \mathcal{L}_A) + \widetilde{S}_{n,\text{logical}}~,
\end{align}
for some state $\widetilde{\rho}_{\text{brane},A}$, where $\widetilde{S}_{n,\text{logical}}$ represents the refined Renyi entropy of the logical state?
If $\widetilde{\rho}_A$ indeed satisfies such a formula, what determines the state $\widetilde{\rho}_{\text{brane},A}$, and why does this state manifest in AdS/CFT as inserting a brane with a particular tension into $\widetilde{\rho}_A$? 

Our main result is to answer these questions. 
In short, yes: one can derive the cosmic brane prescription within the formalism of OQEC, and we do this in Section~\ref{sec:qecrenyi}.
Notably, this derivation requires that the AdS/CFT code has certain special properties, which we now explain by discussing refined Renyi entropy in general OQEC codes.

For a general OQEC code, the refined Renyi entropy of $\widetilde{\rho}_A$ from Eq.~(\ref{eqn:advancedqeccdensitymatrix}) is
\begin{align}
    \widetilde{S}_n(\widetilde{\rho}_A) = \sum_\alpha  p_\alpha^{(n)} \widetilde{S}_n(\chi_\alpha) + \widetilde{S}_{n,\text{logical}}~,
\end{align}
where 
\begin{align}
    p_\alpha^{(n)} = \tr \left( P_\alpha \frac{\widetilde{\rho}_A^n}{\tr(\widetilde{\rho}_A^n)} \right)
\end{align}
and $P_\alpha = \sum_i \ket{\alpha, i}\bra{\alpha,i}$ is a projector onto a particular value of $\alpha$. 
While this equation bears some resemblance to Eq.~(\ref{eqn:idealbranegeneralization}), they do not match in general.

Indeed, there are two aspects of the code that need to be true for Eq.~(\ref{eqn:idealbranegeneralization}) to hold. 
First, it must be the case that 
\begin{align}
    \sum_\alpha p_\alpha^{(n)} \widetilde{S}_n(\chi_\alpha) = \tr\left( \widetilde{\rho}^{(n)}_A \mathcal{L}_A \right)
\end{align}
for some state $\widetilde{\rho}^{(n)}_A$ in the code subspace.
We show that this is true if and only if $\chi_{\alpha}$ is maximally mixed.
Second, when interpreted in the context of the AdS/CFT code, $\widetilde{\rho}_A^{(n)}$ needs to manifest as the state $\widetilde{\rho}_A$ with an inserted cosmic brane of exactly the right tension.  
One of our primary focuses is to demonstrate that CFT states with geometric duals indeed admit such an interpretation, as long as $\chi_{\alpha}$ is maximally mixed.
Formulating this argument requires that we carefully interpret Eq.~(\ref{eqn:advancedqeccdensitymatrix}) in gravity. 
For example, we must understand that each $\alpha$-block corresponds to a particular geometry so that we can interpret some $\alpha$-blocks as geometries with cosmic branes.
We must also understand that CFT states with geometric duals have non-vanishing support $p_\alpha$ on $\alpha$-blocks corresponding to many different classical geometries. 
This way, $\widetilde{\rho}_A^{(n)}$ can have its support predominantly on a different classical geometry than $\widetilde{\rho}_A$ does.
We provide these interpretations in Section~\ref{sec:interpretation}, and we explicitly show how they manifest as a cosmic brane prescription within quantum error-correction in Section~\ref{sec:qecrenyi}.

Also in Section~\ref{sec:qecrenyi}, we emphasize the fact that 
a maximally-mixed $\chi_{\alpha}$ for each $\alpha$ implies both properties needed for a code to match the cosmic brane prescription.
This leads us to conclude that $\chi_{\alpha}$ is maximally-mixed in AdS/CFT. 
This has a number of interesting implications, such as an improved definition of the area operator
\begin{align}
    \mathcal{L}_A = \oplus_\alpha \log \dim (\chi_{\alpha}) \mathbb{1}_{a_{\alpha} \bar{a}_{\alpha}}~.
\end{align}

In Section~\ref{sec:tn}, we discuss the implications of these results for tensor network models of AdS/CFT.
While tensor networks tend to nicely satisfy the RT formula \cite{Swingle:2009bg,Pastawski:2015qua,Hayden:2016cfa}, historically they have struggled to have a non-flat spectrum of Renyi entropies.
Our results suggest that there is a natural way to construct a holographic tensor network that not only has the correct Renyi entropy spectrum, but also computes the Renyi entropies via a method qualitatively similar to the cosmic brane prescription. 

Finally, in Section~\ref{sec:discussion} we conclude with a discussion of implications, future directions and related work. Note that this paper was released jointly with \cite{Dong:2018seb} where similar ideas are discussed.

\section{Operator-algebra Quantum Error Correction}\label{sec:OQEC}
We start by reviewing the framework of operator-algebra quantum error correction (OQEC) as discussed in \cite{Harlow:2016vwg}.
Consider a finite dimensional ``physical" Hilbert space $\mathcal{H}=\mathcal{H}_{A}\otimes \mathcal{H}_{{\bar{A}}}$ and a ``logical" code subspace $\mathcal{H}_{\text{code}}\subseteq \mathcal{H}$.\footnote{More generally the physical Hilbert space need not factorize, e.g. if the boundary theory has gauge constraints. We expect the qualitative features of our result to be unchanged in that case.}
In the context of holography, one can think of $\mathcal{H}$ as the boundary Hilbert space and $\mathcal{H}_{\text{code}}$ as the Hilbert space of the bulk effective field theory (EFT). 

Let $\mathcal{L(H_\text{code})}$ be the algebra of all operators acting on $\mathcal{H}_{\text{code}}$ and $M\subseteq \mathcal{L(H_\text{code})}$ be a subalgebra.
In particular, we require that $M$ be a von Neumann algebra, i.e. it is closed under addition, multiplication, hermitian conjugation and contains all scalar multiples of the identity operator. 

Any von Neumann algebra has an associated decomposition of the Hilbert space given by
\begin{align}\label{eq:decomp}
    \mathcal{H}_{\text{code}}=\oplus_{\alpha} \left( \mathcal{H}_{a_{\alpha}}\otimes \mathcal{H}_{\bar{a}_{\alpha}}\right)~,
\end{align}
such that the operators in the von Neumann algebra are the set of $\alpha$ block diagonal operators that only act non-trivially on the $\mathcal{H}_{a_{\alpha}}$ factor within each block.
Namely, they are of the form
\begin{align}\label{eq:opM}
    \widetilde{O}_M = \oplus_{\alpha} \left(\widetilde{O}_{a_{\alpha}}\otimes \mathbb{1}_{\bar{a}_{\alpha}}\right)~,
\end{align}
where from now onward, we use the ``tilde" to denote objects that naturally act on the code subspace.\footnote{The only exception to this is the notation for the refined Renyi entropy $\widetilde{S}_n$.}
The commutant of $M$, denoted $M'$, is defined by the set of operators that commute with all the operators in $M$.
The operators in $M'$ are then similarly of the form
\begin{align}\label{eq:opM'}
    \widetilde{O}_{M'} = \oplus_{\alpha} \left(\mathbb{1}_{a_{\alpha}}\otimes \widetilde{O}_{\bar{a}_{\alpha}}\right)~.
\end{align}
The center $Z_M$ consists of operators that belong to both $M$ and $M'$ take the form
\begin{align}\label{eq:center}
    \widetilde{O}_{M'} = \oplus_{\alpha} \left(\lambda_{\alpha} \mathbb{1}_{a_{\alpha}}\otimes \mathbb{1}_{\bar{a}_{\alpha}} \right)~,
\end{align}
where $\lambda_{\alpha}$ could in general be different for each $\alpha$ block.

The OQEC code is then defined by requiring that for any state in the code subspace, the operators in the von Neumann algebra $M$ are robust against erasure of the subfactor $\mathcal{H}_{\bar{A}}$ of the physical Hilbert space.
Equivalently, we require that all the operators in $M$ can be represented as physical operators acting non-trivially only on $\mathcal{H}_{{A}}$. In addition, by taking inspiration from AdS/CFT, we restrict to OQEC codes with complementary recovery, i.e. where operators in $M'$ are robust against erasure of $A$.
Thus, we require that for all $\ket{\widetilde{\psi}}\in \mathcal{H}_{\text{code}}$, $\widetilde{O}_M \in M$ and $\widetilde{O}_{M'} \in M'$, there exists $O_A \in \mathcal{L}(\mathcal{H}_A)$ and $O_{\bar{A}} \in \mathcal{L}(\mathcal{H}_{\bar{A}})$ such that
\begin{align}\label{eq:SubDuality}
    \widetilde{O}_M \ket{\widetilde{\psi}} &= O_A \ket{\widetilde{\psi}}\\
    \widetilde{O}_M ^{\dagger} \ket{\widetilde{\psi}} &= O_A^{\dagger} \ket{\widetilde{\psi}}\\
    \widetilde{O}_{M'} \ket{\widetilde{\psi}} &= O_{\bar{A}} \ket{\widetilde{\psi}}\\
    \widetilde{O}_{M'} ^{\dagger} \ket{\widetilde{\psi}} &= O_{\bar{A}}^{\dagger} \ket{\widetilde{\psi}}
\end{align}

Let us pause for a moment to make connections with holography.
Suppose $A$ is a boundary subregion and $\gamma(A)$ is the bulk codimension 2 extremal surface of minimal area anchored to $\partial A$, i.e. the RT surface of $A$.\cite{Hubeny:2007xt,Wall:2012uf}.
The entanglement wedge $\text{EW}(A)$ is defined as the bulk domain of dependence of any bulk spacelike surface $\Sigma$ such that $\partial \Sigma =A\cup \gamma(A)$ \cite{Czech:2012bh,Headrick:2014cta,Jafferis:2014lza}.
Given a pure boundary state, $\gamma(A)=\gamma(\bar{A})$ and thus, $\text{EW}(A)\cup \text{EW}(\bar{A})$ includes a complete Cauchy slice in the bulk.\footnote{If the boundary state is mixed, e.g. a thermal state, one could purify it, e.g. to a thermofield double \cite{Maldacena:2001kr}.}
Interpreting $M$ and $M'$ as the algebra of operators in $\text{EW}(A)$ and $\text{EW}(\bar{A})$ respectively, it is clear that Eq.~(\ref{eq:SubDuality}) is the statement of entanglement wedge reconstruction \cite{Dong:2016eik,Faulkner:2017vdd}.
In holography, the surface $\gamma(A)$ is fixed irrespective of the state of bulk quantum fields at leading order in $G_N$.
In fact even at first subleading order in $G_N$, one could calculate $S(A)$ by keeping $\gamma(A)$ fixed and including bulk entropy corrections at $\mathcal{O}(1)$ \cite{Faulkner:2013ana}.
At higher orders in $G_N$, one has to take into account the quantum extremal surface prescription for $\gamma(A)$ wherein the surface can move depending on the bulk state \cite{Engelhardt:2014gca,Dong:2017xht}.
In general there would be a ``no man's land" region in the bulk that cannot be reconstructed state-independently by either $A$ or $\bar{A}$.
Thus, the OQEC code with complementary recovery only works as a toy model of holography when computing entanglement entropy to $\mathcal{O}(1)$ and hence, all our results hold only to this order.

As shown in \cite{Harlow:2016vwg}, one can equivalently find unitaries $U_A$ and $U_{\bar{A}}$ such that
\begin{align}\label{eq:basis}
    \ket{\widetilde{\alpha,i\,j}}=U_A\,U_{\bar{A}}\left(\ket{\alpha,i}_{A_1^{\alpha}}\ket{\alpha,j}_{\bar{A}_1^{\alpha}}\ket{\chi_{\alpha}}_{A_2^{\alpha}\,\bar{A}_2^{\alpha}}\right)~,
\end{align}
where $\ket{\widetilde{\alpha,i\,j}}\equiv\ket{\widetilde{\alpha,i}}\otimes\ket{\widetilde{\alpha,j}}$ is a complete orthonormal basis for the code subspace.
Here, the Hilbert space $\mathcal{H}_A$ has been decomposed as $\mathcal{H}_A=\oplus_\alpha \left(\mathcal{H}_{A_1^{\alpha}}\otimes\mathcal{H}_{A_2^{\alpha}}\right)\oplus \mathcal{H}_{A_3^{\alpha}}$ where $\dim(\mathcal{H}_{A_1^{\alpha}})=\dim(\mathcal{H}_{a_{\alpha}})$.
A similar decomposition has been applied to $\mathcal{H}_{\bar{A}}$. 

This allows us to write a general density matrix $\widetilde{\rho}$ in the code subspace as
\begin{align}\label{eq:DensMat}
    \widetilde{\rho}=U_A\,U_{\bar{A}}\left(\oplus_\alpha \,p_{\alpha} \, \rho_{A_1^{\alpha}\bar{A}_1^{\alpha}}\otimes \ket{\chi_{\alpha}}\bra{\chi_{\alpha}}_{A_2^{\alpha}\bar{A}_2^{\alpha}}    \right)U^{\dagger}_{\bar{A}}\,U^{\dagger}_A~,
\end{align}
where we choose the normalizations such that $\tr_{A\bar{A}}(\widetilde{\rho})=\tr_{A_1^{\alpha}\bar{A}_1^{\alpha}}(\rho_{A_1^{\alpha}\bar{A}_1^{\alpha}})=\sum_{\alpha}p_{\alpha}=1$.
Restricting to subregion $A$, we obtain
\begin{align}\label{eq:RedDensMat}
    \widetilde{\rho}_A=U_A\left( \oplus_{\alpha} \,p_{\alpha}\, \rho_{\alpha}\otimes \chi_{\alpha} \right) U^{\dagger}_A~,
\end{align}
where we have relaxed the notation by using $\rho_{\alpha}=\tr_{\bar{A}_1^{\alpha}} \left(\rho_{A_1^{\alpha}\bar{A}_1^{\alpha}} \right)$ and $\chi_{\alpha}=\tr_{\bar{A}_2^{\alpha}}\left( \ket{\chi_{\alpha}}\bra{\chi_{\alpha}}_{A_2^{\alpha}\bar{A}_2^{\alpha}}\right)$. 

Using this it is straightforward to compute the von Neumann entropy of $\widetilde{\rho}_A$ and show that it satisfies a Ryu-Takayanagi formula, i.e.
\begin{align}\label{eq:RT}
    S(\widetilde{\rho}_A) = \tr (\widetilde{\rho}\, \mathcal{L}_A)+S(\widetilde{\rho},M)~,
\end{align}
where $\mathcal{L}_A$ is the area operator,\footnote{Note that the important feature of $\mathcal{L}_A$ is that it is localized to the RT surface $\gamma(A)$. For theories of higher derivative gravity, it would naturally correspond to the Dong entropy \cite{Dong:2013qoa}} an operator in the center of $M$ defined by
\begin{align}\label{eq:AreaOp}
    \mathcal{L}_A\equiv \oplus_{\alpha}\,S(\chi_{\alpha}) \mathbb{1}_{a_{\alpha}\, \bar{a}_{\alpha}} ~.
\end{align}
In a gravitational theory with the Einstein-Hilbert action, the area operator is given by 
\begin{align}
    \mathcal{L}_A&=\frac{\mathcal{A}(\gamma(A))}{4G_N}~.
\end{align}
The second term in Eq.~(\ref{eq:RT}) is the algebraic entropy defined by
\begin{align}\label{eq:AlgEntropy}
        S(\widetilde{\rho},M) \equiv -\sum_{\alpha} p_{\alpha} \log p_{\alpha} + \sum_{\alpha} p_\alpha S (\widetilde{\rho}_{a_{\alpha}})~.
\end{align}

\section{Interpretation of OQEC}\label{sec:interpretation}

In order to understand our result in Section~\ref{sec:qecrenyi}, it will be crucial to interpret each piece of the state in Eq.~(\ref{eq:RedDensMat}) in holography.
The unitary $U_A$ is simply a unitary operation that ``encodes" the logical state $\rho_{\alpha}$ by mixing it with the redundant degrees of freedom $\chi_{\alpha}$.
We ignore this piece and focus directly on the bulk reduced density matrix
\begin{align}\label{eq:directsum}
    {\rho}_a =  \oplus_\alpha \,p_\alpha \,\rho_\alpha \otimes \chi_\alpha~,
\end{align}
where one should think of the bulk subregion $a$ as $\text{EW}(A)$.
We interpret these pieces by first reviewing lattice gauge theory, which has a similar block decomposition and was argued in \cite{Donnelly:2016auv} to have a similar interpretation.
We then proceed by analogy for the case of gravity.

\subsection{Lattice Gauge Theory}\label{subsec:LGT}
Understanding the structure of the reduced density matrix in a gauge theory requires dealing with the novel feature that the Hilbert space doesn't factorize across a spatial partition due to gauge constraints \cite{Donnelly:2011hn,Casini:2013rba,Donnelly:2014gva,Donnelly:2016auv}.
This can be easily visualized in lattice gauge theory, where the gauge field lives on the links of the graph, whereas other fields that transform under the gauge symmetry live on the nodes.
These links are necessarily cut when partitioning the vertices to extract the reduced density matrix for a bulk subregion $a$.
A prescription to compute $\rho_a$ was given in \cite{Donnelly:2011hn,Donnelly:2016qqt} and has several consistency checks backing it \cite{Donnelly:2014fua,Donnelly:2015hxa}. 
\begin{figure}[t]
\begin{center}
  \includegraphics[scale=0.9]{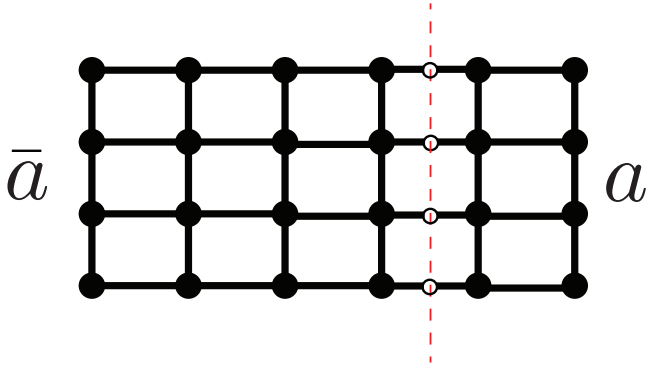} 
  \end{center}
\caption{Decomposing a lattice gauge theory into subregions $a$ and $\bar{a}$ requires the introduction of extra degrees of freedom (denoted as white dots) at the entangling surface (denoted by a dashed red line).}
\label{fig:Lattice}
\end{figure}
In order to find the reduced density matrix $\rho_a$, the prescription is to define an extended Hilbert space by first adding extra degrees of freedom at the entangling surface which are not required to satisfy a gauge constraint (see Figure~\ref{fig:Lattice}). These extra degrees of freedom allow the extended Hilbert space to factorize.
The physical states that satisfy the gauge constraint form a subspace of this extended Hilbert space.

As shown in \cite{Donnelly:2011hn,Donnelly:2014gva}, the requirement that physical states commute with the action of gauge transformations implies that the reduced density matrix must take the form
\begin{align}\label{eq:representations}
    \rho_a = \oplus_{R} \,p(R)\, \rho(R) \otimes \frac{\mathbb{1}_{R}}{\dim R}~,
\end{align}
where the direct sum is over all the different irreducible representations of the entangling surface symmetry group.
Comparing this to Eq.~(\ref{eq:directsum}), the state has a similar form with the restriction that $\chi_{\alpha}=\frac{\mathbb{1}}{\dim \chi_{\alpha}}$.
This $\chi_\alpha$ can be interpreted as the maximally mixed state of the extra degrees of freedom that were added inside the entangling region. 

The block-diagonal structure of Eq.~(\ref{eq:representations}) comes from the following.
The representations $R$ determine all the local gauge invariant observables, e.g. the Casimir operator, and are thus distinguishable within the region $a$.
Hence, the reduced density matrix is in a classical mixture of these superselection sectors with probability distribution $p(R)$. 

It is worth commenting here on one aspect of the connection to gravity. 
When the bulk is treated semiclassically, one might interpret Eq.~(\ref{eq:directsum}) as the gravitational equivalent of Eq.~(\ref{eq:representations}) \cite{Donnelly:2016auv, Lin:2017uzr}. Diffeomorphisms then play the role of the gauge symmetry, and $\chi_\alpha$ represents degrees of freedom added across the boundary to enable factorization  \cite{Donnelly:2016auv}.

\subsection{Gravity}\label{sec:GravInterpret}
Armed with the previous discussion, we now interpret each piece of Eq.~(\ref{eq:directsum}) in gravity, occasionally drawing an analogy to Eq.~(\ref{eq:representations}).

\subsubsection*{$\alpha$-blocks}
Recall that the $\alpha$-blocks were defined by first choosing a von Neumann algebra $M$ and then finding the natural associated block decomposition of the Hilbert space. 
The algebra $M$ has a non-trivial center $Z_M$ since the gauge constraints from diffeomorphism invariance inhibit the low energy bulk Hilbert space from factorizing. 
Then, the simplest physical interpretation of the $\alpha$-blocks is as eigenspaces of the operators in the center $Z_M$.
In holography, these include various gauge invariant observables localized to the RT surface, and in particular the area operator $\mathcal{L}_A$ from Eq.~(\ref{eq:AreaOp}) is one such operator \cite{Donnelly:2016auv}.\footnote{In the presence of other gauge constraints, e.g. U(1) gauge fields in the bulk, there would in general be more center observables, e.g. the electric flux. The $\alpha$-blocks would then be labelled by the values of all such observables. We expect the observables related to non-gravitational constraints to leave the $\alpha$-block structure unchanged at leading order in $G_N$, though this is not important for our analysis.} 
Hence the states in a given $\alpha$-block can be thought of as area eigenstates with the same value of the area operator.

Of course, we do not require infinite precision when comparing eigenvalues. 
Since we compute entropies accurately up to $\mathcal{O}(1)$, we consider two states to be in the same $\alpha$-block if and only if they have the same eigenvalue of $\mathcal{L}_A$ at $\mathcal{O}(1)$.
That said, it is sometimes useful to acknowledge that states with the same eigenvalue at $\mathcal{O}(1/G_N)$ share a classical background geometry.\footnote{Strictly speaking, they need only have the same minimal extremal surface area, but need not be the same geometry. We expect that generically no two different geometries have exactly the same minimal extremal surface area given fixed boundary conditions at infinity. If they do, it would not greatly affect our results.}
For example, empty AdS with a small field excitation might be in a different $\alpha$-block than vacuum AdS because the $\mathcal{A}/4G_N$ differs at $\mathcal{O}(1)$. But the two states still share the same classical background: empty AdS.

Note that our code subspace is relatively large: we include an $\alpha$-block for every geometry with $\mathcal{A}/4G_N$ different at $\mathcal{O}(1)$.
Indeed, we identify the code subspace as the entire Hilbert space of bulk EFT.
This is important for our results in Section~\ref{sec:qecrenyi}, and so we emphasize that we are explicitly assuming this.
We think it is likely to be well-grounded based on an extension of \cite{Almheiri:2016blp}; 
the analysis in that paper motivates the inclusion of different classical background geometries in the code subspace when working at $\mathcal{O}(1/G_N)$.
Once that is established, it is easier to define a code subspace on top of each classical background.
Perhaps an argument along these lines justifies including many classically different geometries in the same code subspace, even when working to $\mathcal{O}(1)$.


We are primarily interested in bulk states that are smooth geometries.
In the $G_N \to 0$ limit, smooth geometries become area eigenstates, and hence have support exclusively on $\alpha$-blocks with one particular value of $\mathcal{A}/4G_N$ at leading order.
When $G_N$ is finite, overlap with other blocks is best computed using the Euclidean path integral \cite{Jafferis:2017tiu}.
This formalism makes it clear that two classically different geometries have $e^{-\mathcal{O}(1/G_N)}$ overlap.
We will go into more depth when we discuss $p_\alpha$ below.

We assume that non-perturbative corrections do not ruin the exactly block-diagonal structure of Eq.~(\ref{eq:directsum}).
As we will see in Section~\ref{sec:qecrenyi}, this suffices to derive a prescription resembling the AdS/CFT calculation.
It is unclear whether this is well-justified, but one argument for it is the following.
We work exclusively in the context of semi-classical gravity, and semi-classical gravity states are non-perturbatively gauge invariant \cite{Jafferis:2017tiu}.
Gauge invariance demands the direct sum structure exactly.
I.e., non-perturbatively small off block-diagonal terms might break non-perturbative gauge invariance.
Again, this argument is intended heuristically, not as a proof. 
For our purposes, we simply assume that the off-diagonal terms are exactly zero.

\subsubsection*{State $\rho_\alpha$}
This physical understanding of $\alpha$-blocks lends itself to an easy interpretation of $\rho_\alpha$.
The state $\rho_\alpha$ is the state of the bulk matter fields when restricted to the subspace of the bulk Hilbert space with a definite value of $\mathcal{A}/4G_N$ to $\mathcal{O}(1)$.
From Eq.~(\ref{eq:AreaOp}), we see that the definite value of $\mathcal{A}/4G_N$ is $S(\chi_\alpha)$.

\subsubsection*{State $\chi_\alpha$}
Since the bulk EFT degrees of freedom are captured by the $\rho_{\alpha}$ state, the $\chi_{\alpha}$ state must correspond to the high energy, quantum gravity degrees of freedom that were integrated out to define the semiclassical gravity EFT.

This is nicely consistent with the fact that the area operator $\mathcal{L}_A$ measures the entropy of these degrees of freedom.
In semiclassical gravity, the generalized entropy of a subregion is defined as
\begin{align}
    S_\text{gen} = \frac{\mathcal{A}}{4G_N} + S_\text{matter}~,
\end{align}
where $S_{\text{matter}}$ is the von Neumann entropy of the bulk matter fields. While $S_\text{gen}$ is defined purely in semiclassical gravity, it is widely believed that it corresponds to the von Neumann entropy of all the degrees of freedom in the ``full" quantum gravity theory.
The entropy of the quantum gravity degrees of freedom that were integrated out shows up as the area term in $S_\text{gen}$. 
Comparing with Eq.~(\ref{eq:RT}), we see that the entropy $S(\widetilde{\rho}_A)$ is interpreted in exactly this way: the part $S(\rho_\alpha)$ is the bulk matter entropy, and $S(\chi_\alpha)$ can be interpreted as the area. 


One can gain further insight into the degrees of freedom described by $\chi_\alpha$ by drawing an analogy to lattice gauge theory.
By comparing Eq.~(\ref{eq:directsum}) and Eq.~(\ref{eq:representations}), one sees that the $\chi_{\alpha}$ degrees of freedom seem analogous to the surface symmetry degrees of freedom, which are in the state $\frac{\mathbb{1}_{R}}{\dim R}$.
This was pointed out in \cite{Donnelly:2016auv,Lin:2017uzr}, in which they argue that because semi-classical gravity is a gauge theory, this should be understood as more than an analogy. One should expect that the $\chi_\alpha$ can be interpreted as playing the role of surface symmetry degrees of freedom for the diffeomorphism group.

This interpretation would come with an interesting implication. Because gauge invariance imposes that surface symmetry degrees of freedom are in the singlet state, it would have to be the case that $\chi_\alpha = \frac{\mathbb{1}_{\alpha}}{\dim \chi_{\alpha}}$. 
Confirming that $\chi_\alpha$ is indeed in this state in the AdS/CFT code will be one of our main results in Section~\ref{sec:qecrenyi}.

\subsubsection*{Distribution $p_\alpha$}
How would we compute $p_\alpha$ within the CFT?
We could first prepare the state $\widetilde{\rho}_A$ with the Euclidean path integral.
Then, based on Eq.~(\ref{eq:RedDensMat}), project onto block $\alpha$
-- which has area eigenvalue $S(\chi_\alpha)$ -- 
and take the trace to isolate $p_\alpha$.

The analogous procedure in the bulk first prepares a bulk density matrix with the Euclidean path integral. 
We define the bulk density matrix as follows. 
Let $(g_-,\phi_-)$ and $(g_+,\phi_+)$ correspond to two classical metric and matter field configurations at Euclidean time $\tau = 0$. 
The density matrix $\rho[g_-,\phi_-;g_+,\phi_+]$ is a functional of these two configurations defined by the path integral: 
\begin{align}\label{eq:bulkpathintegral}
    \rho[g_-,\phi_-;g_+,\phi_+] = \frac{1}{Z}\int_{\substack{ g', \phi'|_\infty = \mathcal{J} \\ (g_-,\phi_-;g_+,\phi_+)}} Dg' D\phi' \,e^{-I_\text{bulk}[g',\phi']}~,
\end{align}
where the notation $(g_-,\phi_-;g_+,\phi_+)$ is taken to enforce the following two boundary conditions, one on the integration over Euclidean time from $(-\infty, 0)$, and the other over $( 0,\infty)$: 
\begin{align}
    &\tau \in (-\infty,0):~~~~~~ g(\tau=0) = g_-~,~~~ \phi(\tau = 0) = \phi_-~~~,\\
    &\tau \in (0,+\infty):~~~~~~ g(\tau=0) = g_+~,~~~ \phi(\tau = 0) = \phi_+~~~.
\end{align}
The other boundary conditions on the path integral come from the AdS asymptotics: $g',\phi'|_\infty$, must be consistent with boundary conditions at infinity set by the boundary sources $\mathcal{J}$.
With this bulk density matrix, one then defines the reduced density matrix $\rho_a$ by tracing out the complement of $a$, which is $\bar{a} \equiv$ EW($\bar{A}$).
This tracing is done by first identifying $g_- = g_+$ and $\phi_- = \phi_+$ in the region being traced out, then integrating over all metric and matter configurations in that region. 

To obtain $p_\alpha$, we wish to project onto states in that $\alpha$-block and then take the trace.
This is performed by tracing out $a$ from $\rho_a$ while including an extra boundary condition in the path integral that the RT area operator takes on definite value $S(\chi_\alpha)$.

All of these bulk path integral computations can be lumped into one step: we compute $p_\alpha$ by performing the entire Euclidean path integral subject to the boundary condition that the $\mathcal{A}/4 G_N$ of $\gamma(A)$ takes on definite value $S(\chi_\alpha)$:\footnote{The path integration must also respect the boundary conditions at infinity.}
\begin{align}\label{eq:palphaintegral}
    p_{\alpha}=\frac{1}{Z} \int_{\substack{\mathcal{L}_A=S(\chi_{\alpha})\\ g',\phi'|_\infty = \mathcal{J}}} Dg'\, D\phi'\, e^{-I_\text{bulk}[g',\phi']}~.
\end{align}

The leading order contribution to $p_\alpha$ can be computed with the saddle-point approximation:
\begin{align}
    p_{\alpha} \approx \frac{e^{-I_\text{bulk}[g_\alpha, \phi_\alpha]}}{Z}\bigg|_{g_\alpha, \phi_\alpha|_\infty = \mathcal{J}}~.
\end{align}
We have denoted the saddle-point metric and field configuration for each $\alpha$ as $g_\alpha,\phi_\alpha$.
We later schematically shorten this to $p_\alpha = e^{-I_\text{bulk}[\alpha]}/Z|_{b.c.}$, for boundary conditions ``b.c." 
Higher order corrections can be computed from a perturbative expansion of the path integral.
We will pay special attention to the leading order piece in Section~\ref{sec:qecrenyi}, because it will play the most important role in connecting to the cosmic brane prescription \cite{Dong:2016fnf} for computing the refined Renyi entropy. 
Indeed, given a set of boundary conditions at infinity, we will interpret some $\alpha$-blocks as geometries with cosmic branes\footnote{Equivalently, we interpret them as geometries with conical deficits.}, and we will see that computing the refined Renyi entropy is exactly like computing the von Neumann entropy of a state with support predominantly in one of these cosmic brane $\alpha$-blocks.

What we have said here is so important to our main point that we emphasize it again now. 
A general CFT state $\widetilde{\rho}$ has non-vanishing support on many $\alpha$, not just the blocks that correspond to its dominant geometric dual. 
This support is computable, and is schematically $p_\alpha = e^{-I_\text{bulk}[\alpha]}/Z$.
Therefore, the boundary reduced density matrix $\widetilde{\rho}_A$ will in general be a mixture of states on different $\alpha$-blocks. 
We will use these facts when arguing that the cosmic brane prescription for computing refined Renyi entropy \cite{Dong:2016fnf} is derivable from within OQEC.

\section{Cosmic Brane Prescription in OQEC}\label{sec:qecrenyi}
Having established the framework, we now use the formalism of OQEC to compute the refined Renyi entropy defined as 
\begin{equation}\label{eqn:Stilde}
    \widetilde{S}_n(\widetilde{\rho}_A) \equiv n^2 \partial_n \left( \frac{n-1}{n} S_n(\widetilde{\rho}_A) \right)~,
\end{equation}
where $S_n(\widetilde{\rho}_A) \equiv \frac{1}{1-n}\log \tr \widetilde{\rho}_A^n$ is the Renyi entropy of subregion $A$. 
In \cite{Dong:2016fnf,Dong:2017xht} it was shown that in AdS/CFT, the refined Renyi entropy of a boundary subregion $A$ is given by
\begin{align}
    \widetilde{S}_n (A) =  \frac{\mathcal{A}}{4 G_N} + \widetilde{S}_n^{\text{bulk}} (a)~,
\end{align}
where $\mathcal{A}$ is the area of a brane with tension $T_n = \frac{n-1}{4 n G_N}$ and extremal area.
The region $a$ is the entanglement wedge of $A$.
Our main result in this section will be to derive this prescription from the formalism of quantum error correction. 
By doing so we uncover an improved understanding of the high energy degrees of freedom in quantum gravity, as well as a more refined definition of the area operator.

\subsection{Quantum Error Correction calculation}\label{sec:MainCalculation}
The refined Renyi entropy of a general density matrix $\rho$ can be shown to satisfy
\begin{align}\label{eqn:modentgeneral}
    \widetilde{S}_n(\rho) = S(\rho^{(n)})=&- \tr\left(\rho^{(n)} \log \rho^{(n)}\right)~,\\
    \rho^{(n)} \equiv& \frac{\rho^n}{\tr(\rho^n)}~.\label{eqn:rhon}
\end{align}
We now use this to compute the refined Renyi entropy of a reduced density matrix in OQEC. 
Consider an arbitrary state $\widetilde{\rho}$.
From Eq.~(\ref{eq:RedDensMat}), we read off the reduced density matrix of subregion $A$ as
\begin{align}\label{eqn:rhoA}
    \widetilde{\rho}_A = U_A \left( \oplus_\alpha p_\alpha \,\rho_\alpha \otimes \chi_\alpha \right) U_A^\dagger~.
\end{align}
Plugging this into Eq.~(\ref{eqn:rhon}) defines the state
\begin{equation}
    \widetilde{\rho}_A^{(n)} = U_A \left( \oplus_\alpha p_\alpha^{(n)} \,\frac{\rho_\alpha^n}{\tr(\rho_\alpha^n)} \otimes \frac{\chi_\alpha^n}{\tr(\chi_\alpha^n)} \right) U_A^\dagger~.
\end{equation}
where 
\begin{align}\label{eqn:pofn}
    p_\alpha^{(n)} \equiv \frac{p_\alpha^n  \tr(\chi_\alpha^n)\tr(\rho_\alpha^n)}{\sum_{\alpha'}p_{\alpha'}^n  \tr(\chi_{\alpha'}^n)\tr(\rho_{\alpha'}^n)}
     = \frac{p_\alpha^n \,e^{-(n-1) S_n(\chi_\alpha) - (n-1) S_n(\rho_\alpha)}}{Z^{(n)}}
\end{align}
represents a normalized probability distribution that depends on $n$.
Using this in Eq.~(\ref{eqn:modentgeneral}) leads us to a crucial ingredient of our main result:
\begin{equation}\label{eqn:mainresult}
    \widetilde{S}_n(\widetilde{\rho}_A) = \sum_\alpha p_\alpha^{(n)} \widetilde{S}_n(\chi_\alpha) - \sum_\alpha p_\alpha^{(n)} \log p_\alpha^{(n)} + \sum_\alpha p_\alpha^{(n)} \widetilde{S}_n(\rho_\alpha)~.
\end{equation}
It is illuminating to note the following connection to the Ryu-Takayanagi formula.
In the special case where $\chi_{\alpha}$ is maximally mixed\footnote{More generally it is given by a normalized projector, which can be thought of as being maximally mixed over the subspace on which it has support.}, one can write
\begin{align}\label{eqn:stilde}
    \widetilde{S}_n(\widetilde{\rho}_A) = \tr \widetilde{\rho}_A^{(n)} \mathcal{L}_A - \sum_\alpha p_\alpha^{(n)} \log p_\alpha^{(n)} + \sum_\alpha p_\alpha^{(n)} S\left(\rho_{\alpha}^{(n)}\right)~.
\end{align}
Written this way, $\widetilde{S}_n(\widetilde{\rho}_A)$ equals the expectation value of the area operator $\mathcal{L}_A$ plus the algebraic von Neumann entropy of the logical degrees of freedom, all evaluated in a state $\widetilde{\rho}^{(n)}_A$ that belongs to the code subspace. 

Notably, to write Eq.~(\ref{eqn:mainresult}) in the form Eq.~(\ref{eqn:stilde}) for arbitrary states $\widetilde{\rho}_A$ in the code subspace, the identification 
\begin{align}
    \sum_\alpha p_\alpha^{(n)} \widetilde{S}_n(\chi_\alpha) = \tr \widetilde{\rho}_A^{(n)} \mathcal{L}_A
\end{align}
must hold term by term, because one could choose states with support on a single $\alpha$-block.
This is equivalent to having $\widetilde{S}_n(\chi_{\alpha}) = S(\chi_{\alpha})$ for all $\alpha$-blocks, and we show in Appendix~\ref{sec:flat} that, for this, it is both necessary and sufficient that $\chi_\alpha$ be maximally mixed.
Moreover, a maximally mixed $\chi_\alpha$ has another important implication.
In the next subsection we shall argue that it is a necessary and sufficient condition for the gravitational interpretation of $\widetilde{\rho}_A^{(n)}$ to be a geometry containing a brane with the precise $n$-dependent tension required to match the cosmic brane prescription in \cite{Dong:2016fnf}.

\subsection{Connection to gravity}\label{sec:GravityConnection}
\paragraph{Review of holographic refined Renyi entropy:}
We start by carefully reviewing the cosmic brane prescription of \cite{Dong:2016fnf}.
One considers a CFT state $\widetilde{\rho}$ that can be prepared by the Euclidean path integral. 
The bulk dual of the reduced density matrix $\rho_A$ is given by the Hartle-Hawking wavefunction \cite{Hartle:1983ai}, which has the saddle-point approximation
\begin{align}\label{eqn:bulkaction}
    \rho_A[g_{-},\phi_{-};g_{+},\phi_{+}] = \frac{e^{-I_\text{bulk}[g,\phi]}}{Z}\bigg|_{\substack{ g, \phi|_\infty = \mathcal{J} \\ (g_-,\phi_-;g_+,\phi_+)}}~,
\end{align}
where $g,\phi$ are the saddle point field configurations given the boundary conditions at Euclidean time $\tau = 0$, $(g_-,\phi_-;g_+,\phi_+)$. 
Moreover, the AdS asymptotics $g,\phi|_\infty$ must be consistent with boundary conditions $\mathcal{J}$ at infinity that define the state $\widetilde{\rho}$.

The cosmic brane prescription computes the refined Renyi entropy $\widetilde{S}_n(\widetilde{\rho}_A)$ in two steps.
First, consider the bulk reduced density matrix
\begin{align}\label{eqn:bulkbraneaction}
    \rho_\text{brane,A}[g_{-},\phi_{-};g_{+},\phi_{+}] = \frac{e^{- n I_\text{bulk}[g,\phi] - (n-1) \frac{\mathcal{A}[g]}{4 G_N}}}{Z}\Bigg|_{\substack{ g, \phi|_\infty = \mathcal{J} \\ (g_-,\phi_-;g_+,\phi_+)}}~.
\end{align}
The action is $n$ times the sum of the bulk action $I_\text{bulk}$ and the area $\mathcal{A}$ of a brane with tension $T_n = \frac{n - 1}{4 n G_N}$ anchored to the boundary of region $A$. 
We refer to this as a ``brane state," because of its bulk interpretation as $\rho$ with an inserted cosmic brane.

Second, in this brane state compute the expectation value of the area operator $\hat{\mathcal{A}}/4G_N$:
\begin{align}
    \widetilde{S}_n(\widetilde{\rho}_A) = \frac{\tr\left(\rho_{\text{brane},A} \hat{\mathcal{A}}\right)}{4 G_N}~.
\end{align}
This computes the refined Renyi entropy to $\mathcal{O}(1/G_N)$. The $\mathcal{O}(1)$ part is computed by including the $\mathcal{O}(1)$ contribution of the area operator and adding the refined Renyi entropy of the entanglement wedge \cite{Dong:2016fnf, Dong:2017xht}.

Let us forestall a possible confusion. It is often said that the refined Renyi entropy equals the area of the extremal cosmic brane, and rightly so. 
Yet, the prescription above was to evaluate the area operator in the brane state. 
The area operator corresponds to the area of the {\it tensionless} extremal surface, so it's not obvious a priori why it should also correspond to the area of the brane with tension. 
In fact, in these brane states, the tensionless extremal surface coincides with the brane.
Indeed, the branes satisfy the very strong condition that their extrinsic curvature tensor vanishes everywhere.\footnote{We thank Aitor Lewkowycz for helping us understand this.} 
(Note, in the limit $n \to 1$ only the trace of the extrinsic curvature remains vanishing.)


This concludes the cosmic brane prescription for computing the refined Renyi entropy.

\paragraph{Branes in quantum error-correction:}
We now argue that this prescription is simply Eq.~(\ref{eqn:stilde}) combined with three special features of gravity. This is our main result. 

The first special feature is the geometric interpretation of $\alpha$-blocks.
As we discussed at length in Section~\ref{sec:GravInterpret}, each $\alpha$ in $\widetilde{\rho}_A$ corresponds to states of definite area eigenvalue.
We are interested in CFT states with smooth geometric bulk duals. 
Such states have support on many $\alpha$-blocks, and different distributions $p_\alpha$ correspond to different smooth geometries.
Of course, one does not expect all possible distributions to correspond to some smooth geometry -- after all, an equal mixture of two different classical geometries is likely not itself a smooth geometry.
The point is that certain mixtures, such as those prepared by the Euclidean path integral, correspond to smooth geometries.
This is crucial for deriving the cosmic brane prescription in OQEC, because immediately below, we will interpret a particular distribution $p_\alpha^{(n)}$ as defining a ``brane geometry''.

The second special feature is the type of support $p_\alpha$ has on many $\alpha$-blocks for smooth geometries.
As we described in Section~\ref{sec:GravInterpret}, $p_\alpha$ can be computed by performing the Euclidean path integral subject to the constraint that the area takes on the appropriate value. 
The leading order in $G_N$ part of $p_\alpha$ is given by the saddle-point approximation, and should be understood as a weight assigned to the classical geometry with that value of the area.
Hence we can write the $p_\alpha$ of the state $\widetilde{\rho}_A$ as 
\begin{align}\label{eqn:gravPalpha}
    p_\alpha = \frac{e^{-I_\text{bulk}[\alpha]}}{Z}\bigg|_{b.c.}~.
\end{align}
Here, the AdS asymptotics are required to match the boundary conditions (b.c.) that define the state $\widetilde{\rho}$. 
With this $p_\alpha$ in hand, plug Eq.~(\ref{eqn:gravPalpha}) into Eq.~(\ref{eqn:pofn}) to obtain the distribution over $\alpha$-blocks of the state $\widetilde{\rho}_A^{(n)}$:
\begin{align}\label{eqn:pn}
    p_\alpha^{(n)} = \frac{e^{-nI_\text{bulk}[\alpha] - (n - 1) S_n(\chi_\alpha) - (n - 1) S_n(\rho_\alpha)}}{Z^{(n)}}\bigg|_{b.c.}~.
\end{align}
Note the boundary conditions are the same as those for $p_\alpha$. 
This is remarkably similar to Eq.~(\ref{eqn:bulkbraneaction}), and includes the known quantum correction to the action from the matter Renyi entropy \cite{Dong:2017xht}.
Indeed, if it were the case that $S_n(\chi_\alpha) = S(\chi_\alpha)$ for all $\alpha$, then $p_\alpha^{(n)}$ would look exactly like $p_\alpha$ but with the action shifted by the area operator:\footnote{We have dropped $S_n(\rho_\alpha)$ for simplicity, but its presence only increases the resemblance.}
\begin{align}\label{eqn:actionQEC}
    I_\text{bulk}[\alpha] \to n I_\text{bulk}[\alpha] + (n-1)\tr(P_{\alpha} \widetilde{\rho}_A P_{\alpha}\mathcal{L}_A)~.
\end{align}
where $P_{\alpha}=\sum_{i} \ket{\alpha,i}\bra{\alpha,i}$ is a projector onto the particular $\alpha$-block we are looking at. Compare this to the AdS/CFT calculation of refined Renyi entropy. 
The action defining the brane state Eq.~(\ref{eqn:bulkbraneaction}) is shifted in exactly this way relative to the state whose refined Renyi entropy we're computing, Eq.~(\ref{eqn:bulkaction}).
This strongly suggests that indeed $S_n(\chi_\alpha) = S(\chi_\alpha)$ for all $\alpha$ and $n$ in the AdS/CFT code.

We already saw separate evidence for this. 
In order to write Eq.~(\ref{eqn:mainresult}) in the form Eq.~(\ref{eqn:stilde}), it is necessary that $\chi_\alpha$ is maximally mixed.
So, the same condition that allows us to relate the refined Renyi entropy to the expectation value of the area operator in a state $\widetilde{\rho}_A^{(n)}$ also guarantees that $\widetilde{\rho}_A^{(n)}$ has an interpretation as the ``brane state'' Eq.~(\ref{eqn:bulkbraneaction})!

This is strong evidence that in the AdS/CFT code, $\chi_\alpha$ is indeed maximally mixed.
We label this the third special feature of gravity. 
With this conclusion, we have completed the argument that Eq.~(\ref{eqn:stilde}) is the cosmic brane prescription. 

It is worth pausing to emphasize where the brane came from, from the point of view of the code. I.e., we now emphasize how one can look at the state $\widetilde{\rho}_A^{(n)}$ and determine that it equals $\widetilde{\rho}_A$ with a brane inserted into the action.
The original sum over $\alpha$ in Eq.~(\ref{eqn:rhoA}) included states of every possible geometry, including geometries with conical deficits. 
Morally, the reweighting $p_\alpha \to p_\alpha^{(n)}$ enhanced the contribution from the conical deficit geometries relative to the others, exactly like inserting a brane. 
It does this via the factor $e^{-(n-1)\tr(P_{\alpha}\widetilde{\rho}_A P_{\alpha}\mathcal{L}_A)}$ in Eq.~(\ref{eqn:pofn}) -- where we have used our conclusion that $S_n(\chi_\alpha) = S(\chi_\alpha)$ to write this in terms of the area operator like in Eq.~(\ref{eqn:actionQEC}).
That factor suppresses geometries with large eigenvalues of the area operator in exactly the same way that inserting $e^{-(n-1)\mathcal{A}/ 4 G_N}$ does when inserted into the bulk action. 

Notably, the fact that $\chi_{\alpha}$ is maximally mixed has a number of interesting implications. 
For instance, it gives us an improved form of the area operator. Instead of Eq.~(\ref{eq:AreaOp}), the AdS/CFT code's area operator is
\begin{align}
    \mathcal{L}_A = \oplus_\alpha \log \dim(\chi_{\alpha}) \mathbb{1}_{a_{\alpha}\, \bar{a}_{\alpha}}~,
\end{align}
where $\dim(\chi_{\alpha})$ is the dimension of the subspace on which $\chi_\alpha$ has support.
This strengthens the argument that $\chi_\alpha$ corresponds to the surface symmetry degrees of freedom in lattice gauge theory. We discuss this more in Section~\ref{sec:discussion}.

Another implication is that states restricted to a single $\alpha$-block have flat Renyi spectra at leading order in $G_N$.
In order for a holographic CFT state to have a non-flat spectrum, it must have support on many $\alpha$-blocks.
In other words, the well-known $n$-dependence of $S_n$ and $\widetilde{S}_n$ for CFTs evidently comes entirely from the $n$-dependent support on $\alpha$-blocks, given by Eq.~(\ref{eqn:pofn}).
This suggests an interesting fix to the notorious inability of tensor networks to have the correct Renyi spectrum \cite{Pastawski:2015qua,Hayden:2016cfa,Qi:2017ohu,Donnelly:2016qqt}. 
We explore this in detail in Section~\ref{sec:tn}.

\section{Tensor Networks}\label{sec:tn}
Holographic tensor networks have modeled certain aspects of AdS/CFT remarkably well \cite{Pastawski:2015qua,Hayden:2016cfa,Qi:2017ohu,Donnelly:2016qqt}. 
Yet, a common mismatch between these models and AdS/CFT has been the flatness of the tensor networks' Renyi spectrum. 
This flatness is in part a result of the maximal entanglement of the bonds. 
Hence some proposals for correcting the spectrum involve modifying the entanglement of the bonds to be less than maximal, often thermal, to match the expected Rindler entanglement across a spatial divide in quantum field theory \cite{Hayden:2016cfa}.

It was proposed in \cite{Almheiri:2016blp} that since the code subspace can be enlarged to include very different background geometries, one might consider a direct sum of tensor networks with different graph structures as a natural toy model for holography.
We now discuss how our results in Section~\ref{sec:qecrenyi} suggest that this approach of defining a ``super tensor network'' (STN) with a Hilbert space that is a direct sum of the Hilbert space of many dissimilar constituent tensor networks resolves the Renyi spectrum mismatch.

The key ingredient is the following. 
Instead of considering a tensor network to be a smooth geometry, tensor networks should be thought of as a single $\alpha$-block! 
I.e., one should think of states on a single tensor network as states with support on a single $\alpha$-block in the code subspace described in Section~\ref{sec:OQEC}.
Indeed, tensor networks are natural area eigenstates, because the ``area'' equals the product of the number of bonds cut and the bond dimension, which is independent of the state.

Moreover, considering tensor networks to be $\alpha$-blocks makes the maximal bond entanglement in tensor networks a feature for matching AdS/CFT's Renyi spectrum, instead of a bug.
If a tensor network's bonds are maximally entangled, the degrees of freedom defining the area operator are maximally mixed (i.e. the state on the bonds cut by the minimal surface is maximally mixed). 
In Section~\ref{sec:qecrenyi}, we called these degrees of freedom $\chi_\alpha$, and indeed we found that for a given $\alpha$, $\chi_\alpha$ is maximally mixed.
Thus, there are strong reasons to treat tensor networks as area eigenstates (i.e. $\alpha$-blocks) and to model smooth geometries as coherent superpositions of tensor networks with different graph structures.
The particular sorts of superpositions that correspond to smooth geometries are discussed in Section~\ref{sec:GravInterpret}.


In fact, tensor networks are even more constraining than a single $\alpha$-block since they are eigenstates of the area operator for arbitrary subregions of the boundary.
In AdS/CFT, it is not precisely clear how one would simultaneously project onto eigenstates of the area operators for different subregions.
RT surfaces anchored to different subregions could cross, and in fact in the time-dependent generalization of RT \cite{Hubeny:2007xt}, generically it wouldn't be possible to constrain all extremal surfaces to lie on the same Cauchy slice.
Since tensor networks represent a coarse grained picture of the bulk with each tensor roughly corresponding to a single AdS volume, it might be possible to impose all the constraints simultaneously.
Since tensor networks manage to perform this simultaneous projection, understanding them better may lead to an improved understanding of holography.
It is also interesting to note that time evolution of tensor networks is not well understood.
This potentially stems from the fact that they model eigenstates of the area operator at a given time and thus, very quickly evolve into states that are not geometric.
It would be interesting to explore this further.

To summarise, a single tensor network with maximal bond entanglement is a good toy model for gravitational states with definite value of the area operator. The code subspace of AdS/CFT is nicely represented by a direct sum over these different tensor networks. 
The tensor network in \cite{Qi:2017ohu} approximately takes this form since there is very small overlap between different networks and thus, any given state is roughly consistent with the direct sum structure.
Our results imply that not only will the STN have a non-flat spectrum, but computing Renyi entropies will be qualitatively similar to the AdS/CFT prescription involving an extremal brane.
One should also compare our results to those of \cite{Han:2017uco}.
They obtained the correct Renyi entropies in the case of $\text{AdS}_3$ by using the specific form of the gravitational action.
We have taken inspiration from the result of \cite{Dong:2016fnf} and instead considered the refined Renyi entropies, which seem to be a more natural quantity in holography. 

\section{Discussion}\label{sec:discussion}

Inspired by the AdS/CFT result that $\widetilde{S}_n$ equals the area of an extremal cosmic brane, we showed that a similar, more general prescription holds for any operator-algebra quantum error correcting code.
This helps better our understanding of the emergent bulk in terms of error correction. 

For a code to satisfy the cosmic brane prescription, the redundant, quantum gravity degrees of freedom split by the entangling surface must be maximally entangled. 
Somehow this fact is encoded in the gravitational path integral, because that was the only input into proving the brane prescription in AdS/CFT.
It would be interesting to see a direct way of proving this within gravity.

We now proceed to discuss some interesting implications of our work and some potential future directions that would be interesting to pursue.

\subsection{Edge Modes and Lattice Gauge Theory}
As we reviewed in Section~\ref{subsec:LGT}, the reduced density matrix in a lattice gauge theory must take the form
\begin{align}
    \rho_A = \oplus_{R} \,p(R)\, \rho(R) \otimes \frac{\mathbb{1}_{R}}{\dim R}~,
\end{align}
where the direct sum is over all the different representations of the entangling surface symmetry group. This strongly resembles states in the OQEC formalism, namely Eq.~(\ref{eq:RedDensMat}), given our conclusion from Section~\ref{sec:qecrenyi} that $\chi_{\alpha}=\frac{\mathbb{1}}{\dim \chi_{\alpha}}$.

There have been various arguments in favour of understanding the bulk as an emergent gauge theory \cite{Harlow:2015lma}. 
The above picture then suggests that the area term in the Ryu Takayanagi formula could be analogous to the $\log \dim R$ term that arises in lattice gauge theory \cite{Lin:2017uzr}. In order for this story to hold true, an important restriction as we saw above was that the state $\chi_{\alpha}$ be maximally mixed. However, we arrived at this from the independent consideration of requiring that the OQEC code satisfy the Dong prescription. Thus, this puts the emergent gravity proposal on a stronger footing.

In order to understand how the Ryu Takayanagi formula arises more precisely, one would have to study the representations of the surface symmetry group in the context of gravity. The results of  \cite{Donnelly:2016auv,Speranza:2017gxd} motivate that the entropy from the $\chi_{\alpha}$ degrees of freedom must scale with the area. The real test would be to obtain the correct prefactor. The results in \cite{Engelhardt:2017aux,Engelhardt:2018kcs,Nomura:2018aus} might help provide a statistical interpretation to understand the edge mode counting in the case of a restricted class of codimension 2 surfaces.

\subsection{Holography in General Spacetimes}
Holography beyond AdS/CFT has been quite elusive as yet. In fact, the AdS/CFT dictionary at length scales shorter than $l_{AdS}$ hasn't been completely understood and has been conjectured to involve the $N^2$ matrix degrees of freedom of the boundary theory in an important way.
Various attempts at understanding holography more generally include \cite{Nomura:2016ikr, Nomura:2017npr, Nomura:2017fyh, Nomura:2018kji,Alishahiha:2004md,Dong:2018cuv,Miyaji:2016mxg,Miyaji:2015fia}.
In each of these cases, there have been attempts to find some form of a Ryu Takayanagi (RT) formula \cite{Sanches:2016sxy,Dong:2018cuv,Miyaji:2015yva}.
To be precise, extremal surfaces anchored to subregions of the proposed boundary theory exist, and satisfy the expected holographic inequalities.
However, it is not clear whether the area of such extremal surfaces really computes the entanglement entropy of some boundary theory.

As \cite{Dong:2016eik,Harlow:2016vwg} showed, if the RT formula is indeed computing the entropy of a boundary subregion, it automatically implies the existence of an error correcting code with subregion duality.
Since our results have demonstrated that any such holographic duality must satisfy Dong's prescription for the Renyi entropy, it should be an additional feature of any of the above proposals in order for them to be consistent.

Another interesting direction to understand sub-AdS locality, while staying within the realm of AdS/CFT, is the $T\bar{T}$ deformation \cite{McGough:2016lol}. 
The $T\bar{T}$ deformation is an irrelevant deformation to the boundary CFT that has been conjectured to be dual to AdS with a Dirichlet boundary at finite radius. 
Interestingly, in \cite{Donnelly:2018bef}, both the Ryu Takayanagi formula and the Dong prescription were shown to work precisely for a very symmetric setup of the $T\bar{T}$ deformation. 
This strongly motivates that one might in fact have a similar error correcting code with subregion duality even without referring to a conformal boundary.

\subsection{Properties of refined Renyi entropy}
Entanglement entropies are known to satisfy various inequalities such as subadditivity and strong subadditivity. Despite the fact that their linear algebra proofs are quite involved, the holographic proofs are remarkably simple \cite{Headrick:2007km,Wall:2012uf}. The geometric dual and the minimization involved in the RT formula easily allow one to prove these inequalities. In fact, a large set of non-trivial inequalities satisfied by holographic states can be proven \cite{Bao:2015bfa,Bao:2015boa}. 

Both Renyi and refined Renyi entropies are not in general subadditive, even though certain interesting classes of states \cite{Camilo:2018css} have been shown to satisfy such inequalities. In the holographic context, since the refined Renyi entropies are also obtained by following a minimization procedure on a dual geometry, one might be led to believe that similar proofs could be used to prove inequalities for the refined Renyi entropies. These could then be used to constrain the class of holographic states, i.e. a holographic refined Renyi entropy cone.\footnote{We thank Ning Bao for discussions about this.}

However, there are subtleties in proving such inequalities due to the back-reaction from the cosmic brane. When considering refined Renyi entropies of two disjoint regions $A$ and $B$, one would in general have to consider different geometries for computing $\widetilde{S}_n(A)$, $\widetilde{S}_n(B)$ and $\widetilde{S}_n(A\cup B)$. 
It's possible that the OQEC formalism provides a useful, alternative way to prove such inequalities.

More generally, understanding properties of the refined Renyi entropy is interesting future work.
The refined Renyi entropies seem to be a more natural generalization of von Neumann entropy in holography than the Renyi entropies. 
They can be computed by a quantity localized to a codimension 2 surface, while the Renyi entropies in general involve a non-local integral, even at leading order in $G_N$. 
It would be interesting to find quantum information theoretic uses for the refined Renyi entropy, perhaps stemming from Eq.~(\ref{eqn:modentgeneral}). 
For example, if the von Neumann entropy of $\rho$ is too difficult to compute experimentally, one could instead compute the $n$-th refined Renyi entropy of $\sigma$, provided $\rho = \sigma^n / \tr(\sigma^n)$.

\subsection{Error Correction and Holography}
Our analysis involved computing a quantity within the quantum error correction framework and comparing it with results from AdS/CFT. This helped us learn about novel aspects of both error correcting codes and AdS/CFT. Thus, it seems like a fruitful direction to analyze other known holographic results within the framework of error correction in order to refine our understanding of the emergent bulk. As we have seen, many results in AdS/CFT, such as the RT formula and the Dong prescription, are simple ``kinematic" results from OQEC. We have also seen that AdS/CFT puts constraints on the type of OQEC that relates the boundary to the bulk. 
It seems plausible that there is much mileage to be gained from simply exploiting the OQEC framework without needing to reference the dynamics of the boundary theory.

A small caveat to keep in mind is that all our results were obtained by working with a finite dimensional Hilbert space. 
Interesting effects might arise from considering infinite dimensional Hilbert spaces. 
For example, \cite{Almheiri:2016blp} found that the Renyi entropies are discontinuous around $n=1$ when taking the large $N$ limit. 
This is also true for the refined Renyi entropies that we have analyzed. 
Some features of infinite dimensional Hilbert spaces could be modeled by using the framework of approximate error correction \cite{Hayden:2018khn}.\footnote{We thank Patrick Hayden for a discussion about this.}
Understanding the nuances of the large $N$ limit is an interesting direction that we leave for future work.


\section*{Acknowledgements}
We thank Ning Bao, Raphael Bousso, Venkatesa Chandrasekaran, William Donnelly, Patrick Hayden, Arvin Shahbazi Moghaddam, Yasunori Nomura, Aron Wall, and especially Stefan Leichenauer and Aitor Lewkowycz for helpful discussions and comments.
We thank Xi Dong, Daniel Harlow and Don Marolf for discussion of their related work \cite{Dong:2018seb} which appeared jointly with this paper.
This work was supported in part by the Berkeley Center for Theoretical Physics, by the National Science Foundation (award number PHY-1521446), and by the U.S. Department of Energy under contract DE-AC02-05CH11231 and award DE-SC0019380.

\appendix
\section{Flat Renyi Spectrum}\label{sec:flat}
Here we prove that if a density matrix $\rho$ satisfies the condition $\widetilde{S}_n (\rho) = S(\rho)$ for all $n$, it must be a normalized projector. We show this by first showing that this condition of having a flat refined Renyi entropy spectrum is equivalent to having a flat Renyi entropy spectrum.

From the definition of the refined Renyi entropy and the above condition, we have
\begin{align}
    \widetilde{S}_n (\rho) &=n^2 \partial_n \left( \frac{n-1}{n} S_n(\rho)\right)\\
    &= S(\rho)~.
\end{align}
We can integrate with respect to $n$ to obtain
\begin{align}
    \int_1^{n'} \frac{S(\rho)}{n^2} &= \left[\frac{n-1}{n} S_n(\rho) \right]_1^{n'}\\
 \implies S(\rho)&=S_{n'}(\rho)~,\label{eqn:flatspectrum}
\end{align}
where this condition is true for arbitrary $n'$. Now we can use the fact that $\rho$ and $\rho^n$ can be simultaneously diagonalized to arrive at the identity
\begin{align}
    \partial_n\,S_n(\rho)=-\frac{1}{(1-n)^2} \sum_{i=1}^{\dim(\rho)} q_i \log\left( \frac{q_i}{p_i} \right)~,
\end{align}
where $p_i$ are the eigenvalues of $\rho$ and $q_i=p_i^n/ (\sum_{i=1}^{\dim(\rho)}p_i ^{n})$. 
If indeed Eq.~(\ref{eqn:flatspectrum}) is true, then the LHS equals zero for all $n$.
For the RHS to equal zero implies the relative entropy (Kullback-Leibler divergence) between probability distributions $q_i$ and $p_i$ vanishes. Using a standard result, we can conclude that the distributions are indeed identical. This gives us our desired result that
\begin{align}
    p_i &= \left(\sum_{i=1}^{\text{rank}(\rho)} p_i ^{n} \right)^{1/(n-1)}\\
    &= \frac{1}{\text{rank}(\rho)}~,
\end{align}
where we have restricted to the non-zero elements only and used the normalization condition. Thus, in its diagonal basis $\rho$ takes the form
\begin{align}
    \rho=\frac{1}{\text{rank}(\rho)}\sum_{i=1}^{\text{rank}(\rho)} \ket{i}\bra{i}~,
\end{align}
namely, it is a normalized projector.
\bibliographystyle{utcaps}
\bibliography{mybibliography}

\end{document}